# Neural Network-derived perfusion maps:

# a Model-free approach to computed tomography perfusion

# in patients with acute ischemic stroke

Umberto A. Gava[a,b], Federico D'Agata[b], Enzo Tartaglione[c], Marco Grangetto[c], Francesca Bertolino[a,b], Ambra Santonocito[b], Edwin Bennink[d,e], Mauro Bergui[a,b]

[a] Neuroradiology Division Molinette Hospital, Turin (Italy); [b] Neuroscience Department, University of Turin (Italy); [c] Informatics Department, University of Turin, Turin (Italy); [d] Department of Radiology, University Medical Center Utrecht, Utrecht, the Netherlands; [e] Image Sciences Institute, University Medical Center Utrecht, Utrecht, the Netherlands.

Corresponding Author: Address correspondence to Umberto A. Gava, Neuroradiology division, Molinette Hospital, Turin (TO), Corso Bramante 88/90, 10126. E-mail: umbertogava@gmail.com

Federico D'Agata (federico.dagata@unito.it), Enzo Tartaglione (enzo.tartaglione@unito.it), Marco Grangetto (marco.grangetto@unito.it), Francesca Bertolino (francesca.bertolino@unito.it), Ambra Santonocito (ambra.santonocito@unito.it), Edwin Bennink (H.E.Bennink-2@umcutrecht.nl), Mauro Bergui (mauro.bergui@unito.it)

# Abstract

Purpose: In this study we investigate whether a Convolutional Neural Network (CNN) can generate clinically relevant parametric maps from CT perfusion data in a clinical setting of patients with acute ischemic stroke.

Methods: Training of the CNN was done on a subset of 100 perfusion data, while 15 samples were used as validation. All the data used for the training/validation of the network and to generate ground truth (GT) maps, using a state-of-the-art deconvolution-algorithm, were previously pre-processed using a standard pipeline. Validation was carried out through manual segmentation of infarct core and penumbra on both CNN-derived maps and GT maps. Concordance among segmented lesions was assessed using the Dice and the Pearson correlation coefficients across lesion volumes.

Results: Mean Dice scores from two different raters and the GT maps were > 0.70 (good-matching). Inter-rater concordance was also high and strong correlation was found between lesion volumes of CNN maps and GT maps (0.99, 0.98).

Conclusion: Our CNN-based approach generated clinically relevant perfusion maps that are comparable to state-of-the-art perfusion analysis methods based on deconvolution of the data. Moreover, the proposed technique requires less information to estimate the ischemic core and thus might allow the development of novel perfusion protocols with lower radiation dose.

Abbreviations:

- Convolutional Neural Network (CNN),
- Cerebral Blood Volume (CBV),
- Cerebral Blood Flow (CBF),
- Time to Peak (TTP),
- CT Perfusion (CTP),
- CT Angiography (CTA),
- Acute ischemic Stroke (AIS),
- Arterial Input Function (AIF),
- Venous Output Function (VOF),
- Impulse Response Function (IRF).

# 1. Introduction

Occlusion of a cerebral artery causes sudden decrease of the blood perfusion in the vascular territory matching the occluded vessel. The peripheral regions of the area affected by the vascular occlusion have their blood flow deficit reduced by the collateral circulation, in comparison to the centre of the affected territory. In fact, ischemic lesions develop rapidly, originating from the centre of the occluded vascular territory and progressively expanding to the most peripheral regions.

From the onset of symptoms, in the ischemic non-functioning area of the brain two different regions may be identified: a central "core", and a peripheral "penumbra", respectively corresponding to areas of irreversible damage and potential recovery, provided recanalization of the occluded vessel. Therefore, identification of core and penumbra may predict the fate of the tissue and drive reperfusion treatments.[1,2]

The extension of core and penumbra may be estimated clinically, from the symptoms and their onset time, or using perfusion techniques, in particular CT Perfusion (CTP). During CTP, a series of low-dose scans are acquired after contrast bolus injection, allowing to compute time-density curves; deconvolution of these curves allows for generating parametric maps to track perfusion parameters dynamics. Cerebral Blood Volume (CBV), Cerebral Blood Flow (CBF), time to peak (TTP) and mean transit time (MTT) are the estimated perfusion parameters most frequently used in clinical practice. CBV is considered a marker of core, while CBF, TTP, MTT mark penumbra.[3–5]

The role of CTP is particularly critical in patients with unknown time from onset, or out of the 4.5 and 6 hours windows used to select patients for intra-venous and intra-arterial reperfusion treatments, respectively. Two trials (DAWN and DEFUSE) demonstrated clinical usefulness of intra-arterial reperfusion in patients selected using CBV and CBF estimation based on CTP.[6,7]

Different algorithms are used to perform deconvolution of time-intensity curves, some of which are not public and may produce largely different maps.[8] In an ideal setting of limited noise, limited variance and no movement artefact, a pixel-by-pixel analysis, as performed by deconvolution-based algorithms, is probably the best choice to obtain realistic, affordable and reproducible maps. Unfortunately, in the real world, information has to be kept redundant in order to overcome problems due to noise, large variances and movement artefacts. In practice, this imposes to obtain more slices, requiring a larger number of acquisitions and more X-ray exposure for the patients, the extraction of an arterial input function (AIF), and a series of spatial pre-processing steps for noise and variance reduction.

Luckily, employing Machine Learning approaches to the problem of deconvolving time-intensity curves, offers several potential advantages over canonical algorithms, allowing the extraction of information that is relatively insensitive to noise, misalignments and variance.

In our study we want to explore whether a properly trained Convolutional Neural Network (CNN), based on a U-Net-like structure, can generate clinically relevant parametric maps of CBV, CBF and time to peak TTP on a pre-processed dataset of CTP images. CTP images were obtained from a real-world dataset of patients with acute ischemic stroke (AIS), no large lesions on non-contrast CT scan, and candidates for reperfusion therapies.[9] This dataset was chosen because it corresponds to the one used in the DAWN and DEFUSE trials, for which perfusion studies drive reperfusion therapies.

## 2. Materials and Methods

### 2.1 Clinical Data

Perfusion data from 127 consecutive patients were retrospectively obtained from hospital PACS. 12 of them, with unreliable data, were discharged (no contrast seen, excess of movements premature termination of acquisition). The remaining 115 datasets were randomly split into a subset of 100, to train the CNN, and 15, to validate results.

CTP acquisition parameters were as follows: Scanner GE 64 slice, 80 kV, 150 mAs, 44.5 sec duration, 89 volumes (40 mm axial coverage), injection of 40 ml of Iodine contrast agent (300 mg/ml) at 4 ml/s speed.

### 2.2 Calculation of ground truth maps

We calculated perfusion maps, including CBF, CBV, TTP, using a standard pipeline of spatial pre-processing and a state-of-the-art fast model-based non-linear regression (NLR) method developed By Bennink e al.[10]

Motion correction was done using a rigid registration method and subsequently all images were filtered implementing a bilateral filter.[11,12]

AIF and Venous output function (VOF) calculation was done automatically on a 100 voxels sample.

The box-shaped model developed by Bennink et al. describes the impulse response function (IRF) of the perfused tissue in terms of CBV, MTT and tracer delay. The box-shaped IRF enables fast NLR analysis, which is critical in a clinical setting such as ischemic stroke.

Time attenuation curve of the tissue and the relative CBV, CBF and TTP maps are estimated using the calculated AIF with the computed IRF.[10]

2.3 CNN training

In this section we are going to provide an overview on the methods used to train the artificial neural network model on perfusion CT scans in order to obtain CBV, CBF and TTP maps as outputs.

The filtered registered images are the only input provided to the trained U-Net like architecture.[13] This particular neural network architecture has been originally developed for image segmentation: however, it proved to be effective also to solve other tasks.[14,15] The only proposed change here is the use of average pooling layers in place of max pool: the use of the standard max-pool layer in our context is a suboptimal since we do not expect sparse features to be extracted. In our case, we aim at using it for perfusion map inference: instead of using standard cross entropy loss, or dice score/focal loss, which are typical for training segmentation tasks, we minimize the mean squared error loss (MSE), which is compatible to the desired ground truth output. No additional information (like the AIF) was provided to the CNN: all the information is implicitly extracted or inferred from the registered CT scans.

The CT scans are processed as a 3D tensor, where the third dimension is represented by the time scale. Hence, depending on the chosen time granularity, the number of input channels changes accordingly.

The CNN structure is displayed in figure 1.

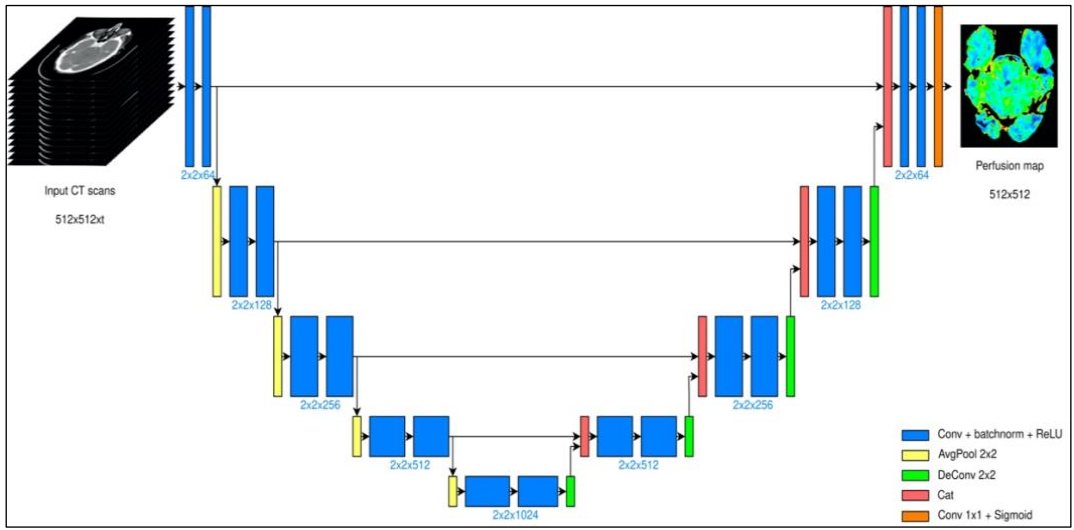

Figure 1. U-Net architecture deployed. The model takes scans of size 512x512, acquired in t different time instants. The bottleneck layer is placed after four encoding stages, and the output is a 512x512 map

The output of the model is a 512x512 map (figure 2), where all the pixel values are normalized in the range [0; 1]. The entire model has been trained using SGD optimization strategy with learning rate decay policy self-tuned according to the performance on the validation set.

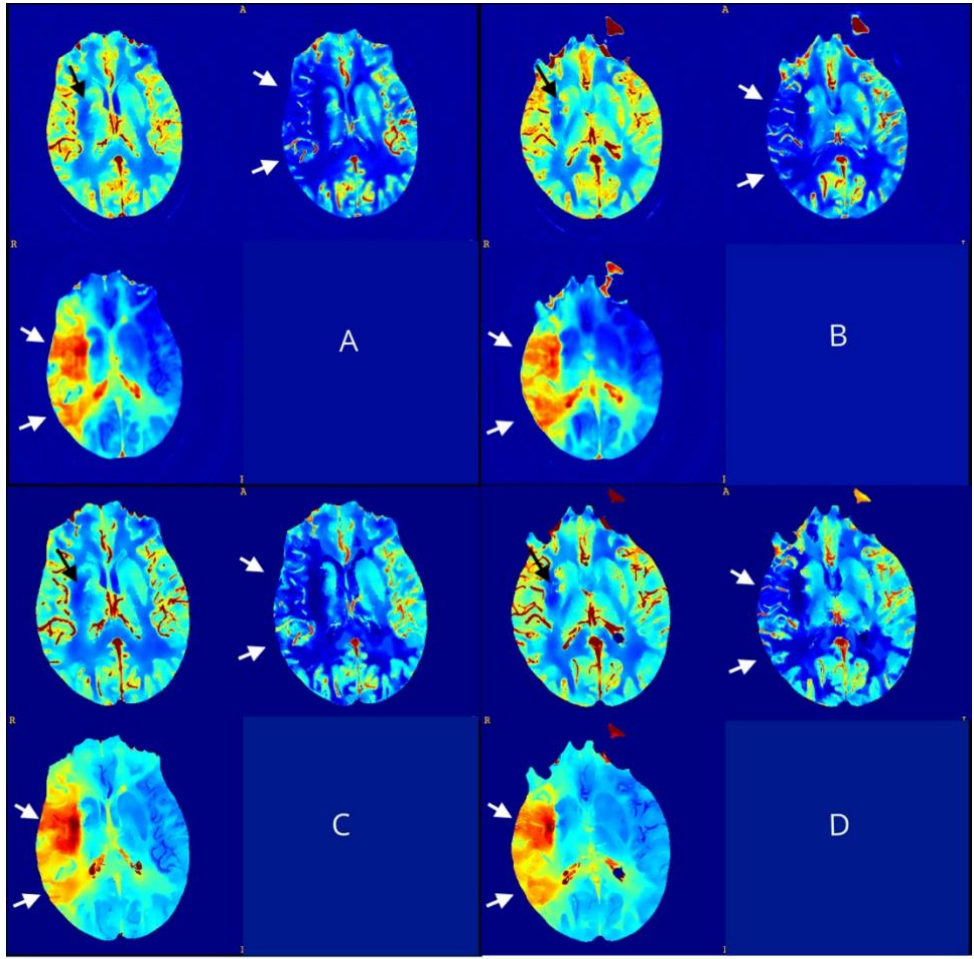

Figure 2. (A, B) CNN outputs maps from a validation set (CBV, CBF and TTP); (C, D) matching sections of GT maps. There is a small infarct core displayed in the CBV map at the right basal ganglia (black arrows) and an extended penumbra showed in the CBF and TTP maps across right middle cerebral artery territories (white arrows).

## 2.3 CNN Validation

Validation is carried out through manual segmentation of the infarct core (CBV) and penumbra (CBF, TTP) on 15 sets of CNN-parametric maps by two expert radiologists using ITK-SNAP opensource software.[16] Segmentation is carried out section-wise following the axial direction.

To avoid bias induced from repetitive evaluation of the same patient, GT-maps segmentation is sorted out the same way by a third Radiologist.

An example of core segmentation on both CNN and GT maps is displayed in figure 3.

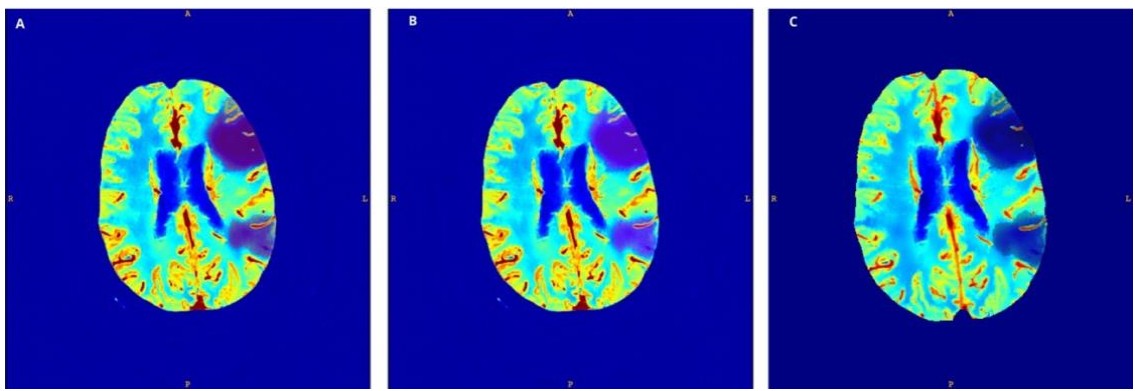

Figure 3. core segmented on CNN CBV map by rater 1 (A) and rater 2 (B); core segmented on GT map (C)

The CNN segmented volumes from both raters were matched with the GT to assess overlapping regions by calculating the Dice Similarity Coefficient (DSC). DSC > 0.70 was considered as good-matching.

DSC were also calculated matching CNN segmentations from different raters to evaluate inter-rater concordance.

Pearson correlation coefficient (r) was used to assess the relationship between the lesion volume on GT and CNN maps. Statistical analysis was performed using the 3D-convert toolbox from ITK-SNAP and SPSS software.

## 3. Results

3 out of 15 CTP datasets used as CNN validation exhibited normal perfusion parameters on both GT and CNN-parametric maps and resulted negative for vessel occlusion on CT Angiography (CTA). Normal perfusion maps were excluded from DSC analysis to avoid overestimation of the segmentations comparison results. Segmented core (CBV) and hypo-perfused regions (CBF/TTP) volumes of the remaining patients are shown in figure 4 and 5; 3 out of 12 patients presented with only penumbra territories without ischemic cores.

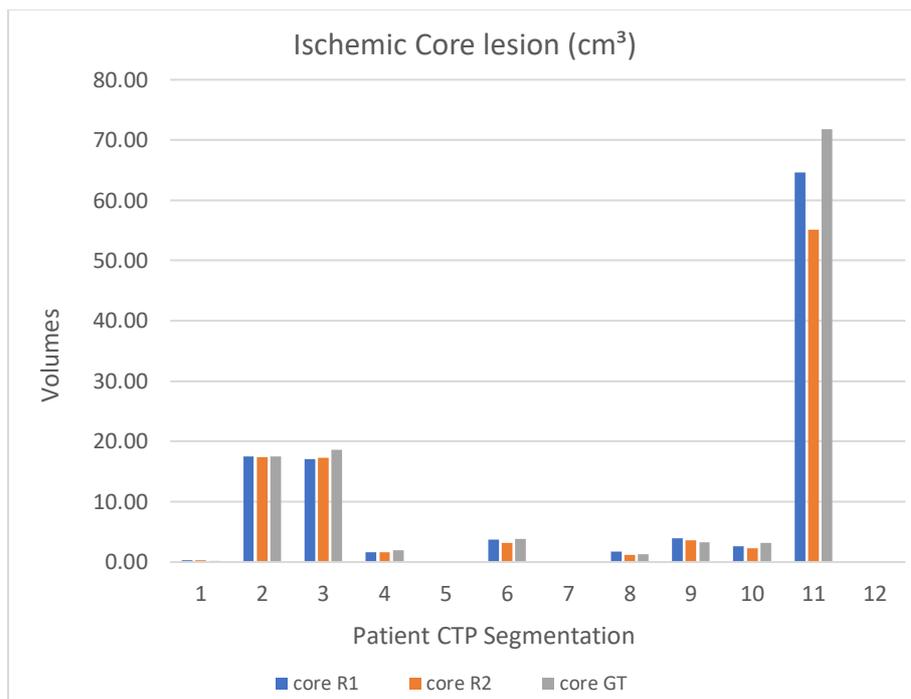

Figure 4. Test sample Ischemic Core segmentation volumes from Rater1 (R1), Rater2 (R2) and GT

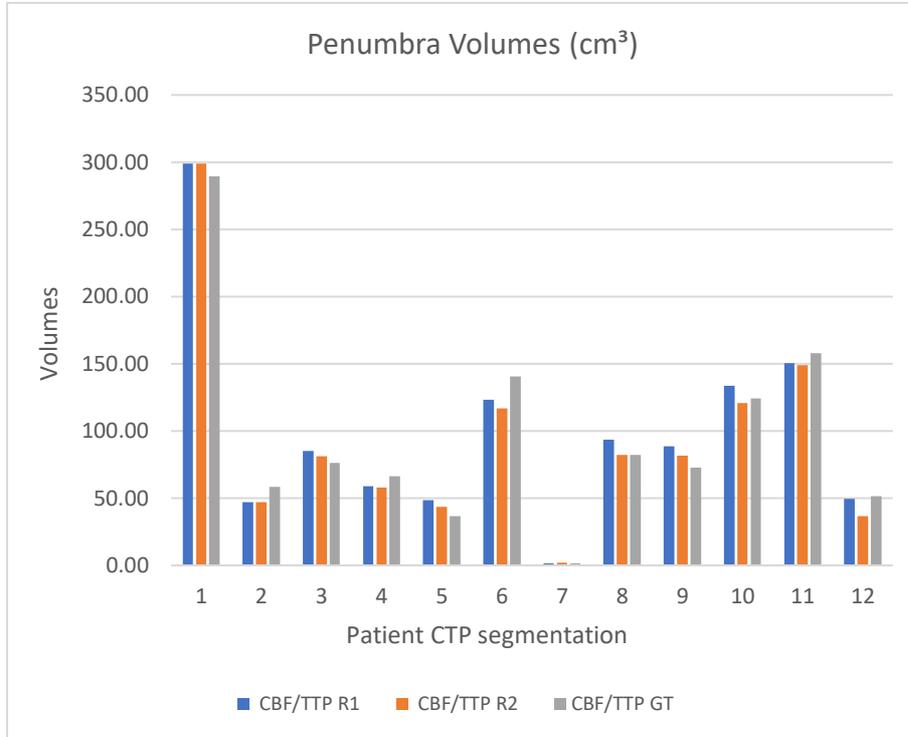

Figure 5. Test sample CBF/TTP segmentation volumes from Rater1 (R1), Rater2 (R2) and GT

Mean DSC for all CBV lesions and CBF/TTP lesions is > 0.70 (Rater1 DSC-CBV 0.82 range 0.71-0.95; Rater2 DSC-CBV 0.79 range 0.73-0.88; Rater1 DSC-CBF/TTP 0.85 range 0.65-0.93; Rater2 DSC-CBF/TTP 0.83 range 0.60-0.92). DSC resulting from segmentation matching are presented as mean and standard deviation (SD) in table 1.

We also find a strong positive correlation (r= 0.99, r= 0.98 with p<0.001) between CBV - CBF/TPP lesion volume on GT and CNN maps for both raters (table 1).

Table 1. Average DSC and lesion volumes Pearson correlations

|  | CBV | CBF/TTP |
|---|---|---|
| Rater1/GT (DSC) | 0.82 ± 0.07 | 0.85 ± 0.07 |
| Rater2/GT (DSC) | 0.79 ± 0.05 | 0.83 ± 0.09 |
| Rater1/Rater2 (DSC) | 0.85 ± 0.10 | 0.87 ± 0.06 |
| Rater1/GT Pearson r | 0.99 | 0.98 |
| Rater2/GT Pearson r | 0.99 | 0.99 |

Table 1. Resulting DSC of CBV and CBF/TTP segmentation expressed as mean and SD; Pearson correlation between CBV-CBF/TTP volumes on GT and CNN maps

# 4. Discussion

The study demonstrates that parametric maps generated by our CNN can compete with a state-of-the-art CTP NLR algorithm, when working on pre-processed images. There is a high DSC and strong linear correlation found between CNN and GT segmented volumes.

The performance of our CNN in estimating ischemic core and penumbra is comparable to a state-of-the-art CTP NLR algorithm.

## 4.1 AIF selection is redundant

To precisely estimate perfusion parameters, the proposed CNN requires only registered CT scans while deconvolution-based CT and MRI brain perfusion analysis methods need additional inputs, such as the AIF curve measured in a large feeding artery.[17] This suggests that our CNN is capable of combining information from arterial and tissue signals to obtain quantitative estimates for the CBV, CBF and MTT. Differently, other recent automated CTP analysis methods such as RAPID,[18] use AIF and the venous output function from a major venous system before computing perfusion parameters and subsequently estimating ischemic core and penumbra.[19]

## 4.2 Translation to a general population

Our validation and training CTP datasets are obtained from a population of patients eligible for reperfusion therapy, with limited or no hypodense lesions on non-contrast CT.
Such a population was targeted to simulate the clinical setting where CTP is a key parameter for clinical decisions: when time of onset is not known, CTP allows to effectively select patients for treatment, as shown in DAWN and DEFUSE-3 trials.[6,7] Furthermore, our method shows accurate performance in patients with no vessel occlusion on CTA and normal CTP parameters on GT maps, therefore suggesting that our CNN approach should yield reliable results even within the general population. In order to confirm this hypothesis, though, validation on a wider sample of CTP data is needed. Moreover, hypodense lesions may mark ischemic core on CTP; this introduces additional information that may easily be exploited by CNNs without estimating the tissue-curve offset, as required, instead, by deconvolution-based algorithms.

### 4.3 Limitations

Our CNN was tested on a high dose/SNR dataset with limited axial coverage, thus an application to noisier datasets is required for a complete comparison with state-of-the-art techniques. Moreover, the CNN should be re-trained using the latest protocols, with extended axial coverage, from different scanners in order to confirm our results while working on different environments.

### 4.4 Applications and future developments

Considering the apparently similar performance of CNNs and deconvolution-based algorithms, one might ask why the former approach might be preferable. Although machine learning algorithms have largely proven to overcome conventional image processing algorithms in practically every field, applications to CTP imaging are still limited (segmentation,[20–22] noise reduction,[22–24] novelty detection,[23,25] radiation dose reduction[23]). In particular, up until now generation of synthetic maps has been done only with MRI DSC perfusion by Ho et al. and Meier et al.[26,27] Meier et al obtained results similar to ours: they compared the performance of a commercial FDA-approved perfusion software and a CNN not only to generate T-Max MRI perfusion maps, but also to identify selection criteria for reperfusion therapies. They concluded that CNN-based approaches may lead to a greater standardization, faster analysis pipeline and increased robustness.[26] Ho et al., instead, estimated voxel-wise MRI perfusion parameters using a deep learning approach exploiting the concentration time curve and AIF as inputs.[27] Their approach, however, proved to be time consuming and thus not ideal for clinical practice.

In this preliminary work, pre-processed images were used as input. Further developments include processing of raw CT acquisitions directly: as demonstrated by novel deep learning models capable of recovering high quality CT images from low dose CT scans of perfusion protocols,[24] a certain amount of information is redundant in CTP images, and currently neural-network-based algorithms probably represent the most effective way to extract it.

Moreover, ischemic core estimate on CNN-derived maps could be compared with more sensitive gold standards such as MRI diffusion imaging, to obtain further information on how our method performs in clinical environment.

## Conclusions

The proposed CNN generated informative perfusion maps of patients with AIS, fairly approximating perfusion mismatch in brain tissues. Our model can fairly compete with state-of-the-art perfusion analysis methods in estimating CBF, CBV and TTP. Moreover, the growing potential of machine learning-based methods of perfusion analysis can lead to new improved standards in terms of acquisition protocols and deployed radiation dose.

## Disclosure of interest

The authors declare that they have no competing interest.

## Acknowledgements

This project was supported by the European Union's Horizon 2020 research and innovation programme under grant agreement No 825111, DeepHealth Project. Giacomo Vaudano MD, Andrea Boghi MD from the Neuroradiology Department of San Giovanni Bosco Hospital (Turin) and Simona Veglia MD from the Radiology Department of Molinette Hospital (Turin) provided help gathering the training dataset. Maarten Terpstra PhD discussed with us the topic and helped setting up the project.